**Stacking polymorphism in PtSe$_2$ drastically affects its electromechanical properties**

*Roman Kempt, Sebastian Lukas, Oliver Hartwig, Maximilian Prechtl, Agnieszka Kuc, Thomas Brumme, Sha Li, Daniel Neumaier, Max C. Lemme, Georg Duesberg, Thomas Heine\**

Roman Kempt, Dr. Thomas Brumme, Prof. Dr. Thomas Heine
Chair of Theoretical Chemistry, Technische Universität Dresden
Bergstrasse 66, 01069 Dresden, Germany
E-mail: thomas.heine@tu-dresden.de

Dr. Agnieszka Kuc, Prof. Dr. Thomas Heine
Helmholtz-Zentrum Dresden-Rossendorf
Permoserstrasse 15, 04318 Leipzig, Germany

Prof. Dr. Thomas Heine
Department of Chemistry
Yonsei University, Seodaemun-gu, Seoul 120-749, Republic of Korea

Sebastian Lukas
Chair of Electronic Devices, RWTH Aachen University, Otto-Blumenthal-Str. 2, 52074 Aachen, Germany

Dr. Sha Li
AMO GmbH, Advanced Microelectronic Center Aachen, Otto-Blumenthal-Str. 25, 52074 Aachen, Germany

Prof. Dr. Daniel Neumaier
Chair of Smart Sensor Systems, Bergische Universität Wuppertal, Lise-Meitner-Str. 13, 42119 Wuppertal, Germany
AMO GmbH, Advanced Microelectronic Center Aachen, Otto-Blumenthal-Str. 25, 52074 Aachen, Germany

Prof. Dr. Max C. Lemme
Chair of Electronic Devices, RWTH Aachen University, Otto-Blumenthal-Str. 2, 52074 Aachen, Germany
AMO GmbH, Advanced Microelectronic Center Aachen, Otto-Blumenthal-Str. 25, 52074 Aachen, Germany

Oliver Hartwig, Maximilian Prechtl, Prof. Dr. Georg S. Duesberg
Insitute of Physics, Faculty of Electrical Engineering and Information Technology (EIT 2), Universität der Bundeswehr München, Werner-Heisenberg-Weg 39, 85577 Neubiberg, Germany

Keywords: PtSe$_2$, two-dimensional materials, piezoresistive sensor, Raman characterization, stacking disorder, density-functional theory

**Abstract**

PtSe$_2$ is one of the most promising materials for the next generation of piezoresistive sensors. However, the large-scale synthesis of homogeneous thin films with reproducible electromechanical properties is challenging due to polycrystallinity. We show that stacking phases other than the 1T phase become thermodynamically available at elevated temperatures that are common during synthesis. We show that these phases can make up a significant fraction in a polycrystalline thin film and discuss methods to characterize them, including their Seebeck coefficients. Lastly, we estimate their gauge factors, which vary strongly and heavily impact the performance of a nanoelectromechanical device.

**1. Introduction**

Two-dimensional (2D) materials are excellent candidates for next-generation nanoelectromechanical devices.[1] They feature high in-plane stiffness and strength but are easily bend,[2] for example as suspended membranes spanned over a cavity.[1] Their electrical response can be tailored with their thickness,[3–5] including metal-to-semiconductor transitions for a reduced number of layers.[6,7] Especially, the noble-metal dichalcogenides (NMDCs), such as PtSe$_2$,[8] excel in piezoresistive sensors due to their high gauge factors,[9,10] long-term stability at ambient conditions,[10,11] and low-temperature synthesis.[12] Furthermore, they have successfully been applied in optical devices, such as phototransistors[13,14] and photodetectors.[11,15–17]

For integrated devices, chemical vapor deposition (CVD) and thermally assisted conversion (TAC) are the preferred options to obtain large-scale thin films of high-quality NMDCs with controllable thicknesses.[1,10,18] PtSe$_2$ is especially promising in this regard because of the low temperature of 400 °C needed for TAC, which is compatible with complementary metal-oxide semiconductor (CMOS) integration.[1,9,12,19] In both cases, challenges arise due to the polycrystallinity of such-obtained films (including polymorphism[20,21] and random crystallite

alignment[22]), the role of the substrate, as well as contacting them in the CMOS integration process.[1] The TAC process has the advantage that additional substrate transfer steps may be avoided by area-selective growth.[23] At every step, a non-invasive and timely characterization of the films is required, which is typically carried out using Raman spectroscopy.[20]

Previously, we performed an extensive study of different NMDC polytypes with respect to the metal coordination, e.g., trigonal prismatic coordination in the $MoS_2$-type 2H phase and octahedral coordination in the $CdI_2$-type 1T phase.[24] In this nomenclature, the number indicates the number of layers per unit cell and the letter abbreviates the crystal system.[25] $PtSe_2$ has been observed in different coordination phases: For instance, Wang et al.[26] showed the formation of $MoS_2$-like 2H-$PtSe_2$ nanoflakes by chemical vapor deposition at 900 °C, while Tong et al.[27] reported that single-layer of the same material was formed by liquid immersion of Pt(111) in $Na_2Se$. Furthermore, Lin et al.[21] observed 1H/1T in-plane $PtSe_2$ heterostructures at different annealing temperatures, while Lei et al.[28] obtained trigonal prismatic bilayer $PtSe_2$. The octrahedrally-coordinated 1T phase of $PtSe_2$ is regarded as the most stable phase.[29] It attracted great attention, because it is semimetallic in the bulk, but becomes semiconducting for fewer layers.[6,7,30] However, other stacking phases of the 1T phase have been investigated to much less extent. To clearly distinguish the metal environments, in the following, we introduce the superscript *O* for octahedral coordination and the superscript *T* for trigonal prismatic coordination. A figure illustrating this naming scheme is given in **Figure S1**. All stacking phases in this study feature octahedral coordination.

A rhombohedral stacking phase with three layers per unit cell ($3R^O$-$PtSe_2$) has been observed as a minor side phase by O'Brien et al.[20] and has been studied by Villaos et al.[31] using DFT simulations. They conclude that it is close in energy to the $1T^O$ phase at 0 K and semiconducting in the bulk. Since other stacking orders of the $1T^O$ phase appear closer in energy than other coordination phases, they can grow competitively at elevated temperatures in TAC and CVD processes, contributing to the formation of polycrystalline thin films. Importantly, the symmetry reduction from the high-symmetry 1T$^O$ phase to lower-symmetry stacking phases can lead to semiconducting properties. This has a significant impact on the electronic characteristics of polycrystalline thin films: For example, other stacking phases might help to explain the large discrepancy of electronic mobilities of PtSe$_2$ found in literature, ranging from lower than 1 cm$^2$ V$^{-1}$ s$^{-1}$ [19,32,33] to 625 cm$^2$ V$^{-1}$ s$^{-1}$.[22]

In this work, we performed an intensive study of the stacking phases of layered PtSe$_2$ using density-functional theory (DFT) to characterize their role in the formation of polycrystalline films in the temperature range between 0 and 1000 K. Notably, we find that lower-symmetry stacking orders than the previously reported AA-stacking order in 1T$^O$-PtSe$_2$[20,29] can make up a significant fraction at elevated temperatures. We confirm the formation of the 3R$^O$ phase reported by Villaos et al.[31] at experimental temperatures, as well as four other metastable stacking phases (2H$^O$, 3T$^O$, 6R$^O$ and 3A$^O$, where A stands for anorthic to avoid confusion between trigonal and triclinic). These have significant impact on the electronic properties of PtSe$_2$ thin films, including semiconducting behavior, large gauge factors, and anisotropic conductivities. We show that the stacking phases cannot be easily distinguished by their Raman signatures and would require HRTEM analysis. We show that the correlation to thermoelectric properties, such as the Seebeck coefficient, may yield a good indication of the presence of other stacking phases.

## 2. Results and Discussion

### 2.1. Structures, Stabilities and Characterization

We arrive at six thermodynamically likely stacking phases ($1T^O$, $2H^O$, $3R^O$, $3T^O$, $6R^O$, $3A^O$) shown in **Figure 1a and 1b** by sampling above two hundred possible stacking orders (details in the **Supporting Information**). They are summarized in **Table 1**.

**Table 1.** Nomenclature of the six stacking orders obtained in this work, their electronic band gaps ($\Delta$) of bulk forms (calculated at HSE06+SOC level), their average interlayer distances ($d$), and estimated "in-plane" gauge factors (GF) and "randomly-aligned" gauge factors (GF*) at the PBE level.

| label | stacking | crystal family | space group | $\Delta$ (eV) | $d$ (Å) | GF | GF* |
|---|---|---|---|---|---|---|---|
| $1T^O$ | AA | trigonal | $P\bar{3}m1$ | 0.00 | 4.957 | 6 to 10 | 1 to 4 |
| $2H^O$ | AB | hexagonal | $P6_3mc$ | 0.00 | 5.385 | -1 to 12 | -4 to 7 |
| $3R^O$ | ABC | rhombohedral | $R\bar{3}m$ | 0.63 | 5.703 | -43 to 16 | -34 to 10 |
| $3T^O$ | AAB | trigonal | $P3m1$ | 0.00 | 5.242 | 5 to 9 | -24 to 12 |
| $3A^O$ | ABC | anorthic[a] | $P1$ | 0.44 | 5.496 | -4 to 22 | -4 to 3 |
| $6R^O$ | AABBCC | rhombohedral | $R\bar{3}m$ | 0.00 | 5.326 | -346 to -63 | -370 to -92 |

[a]We label this stacking order as anorthic instead of triclinic to avoid confusion with trigonal $1T^O$.

All six stacking orders are locally stable showing no imaginary phonon modes (see **Figure S2**). The $1T^O$, $2H^O$, $3T^O$, and $6R^O$ stacking phases are bulk semimetals (see **Figure 2a** and **Figure S3**), while $3R^O$ and $3A^O$ are semiconductors. The $3R^O$ and $3A^O$ stacking phases have much larger mean interlayer distances than the $1T^O$ phase (see **Figures S4 and S5**). This may indicate that a high-pressure synthesis favors the formation of the $1T^O$ phase with smaller interlayer distance. In the $3T^O$, $3A^O$ and $6R^O$ phases, the interlayer distance disproportionates between different stacking regions (see **Figure S5**). Importantly, regions with the same stacking (e.g., AA, BB, CC) always feature the smallest interlayer distance (4.957 Å to 5.032 Å). Differently stacked regions feature interlayer distances between 5.385 Å to 5.703 Å. This may lead to additional X-ray reflections. For comparison, we show the simulated powder X-ray diffraction patterns in **Figure S6**.

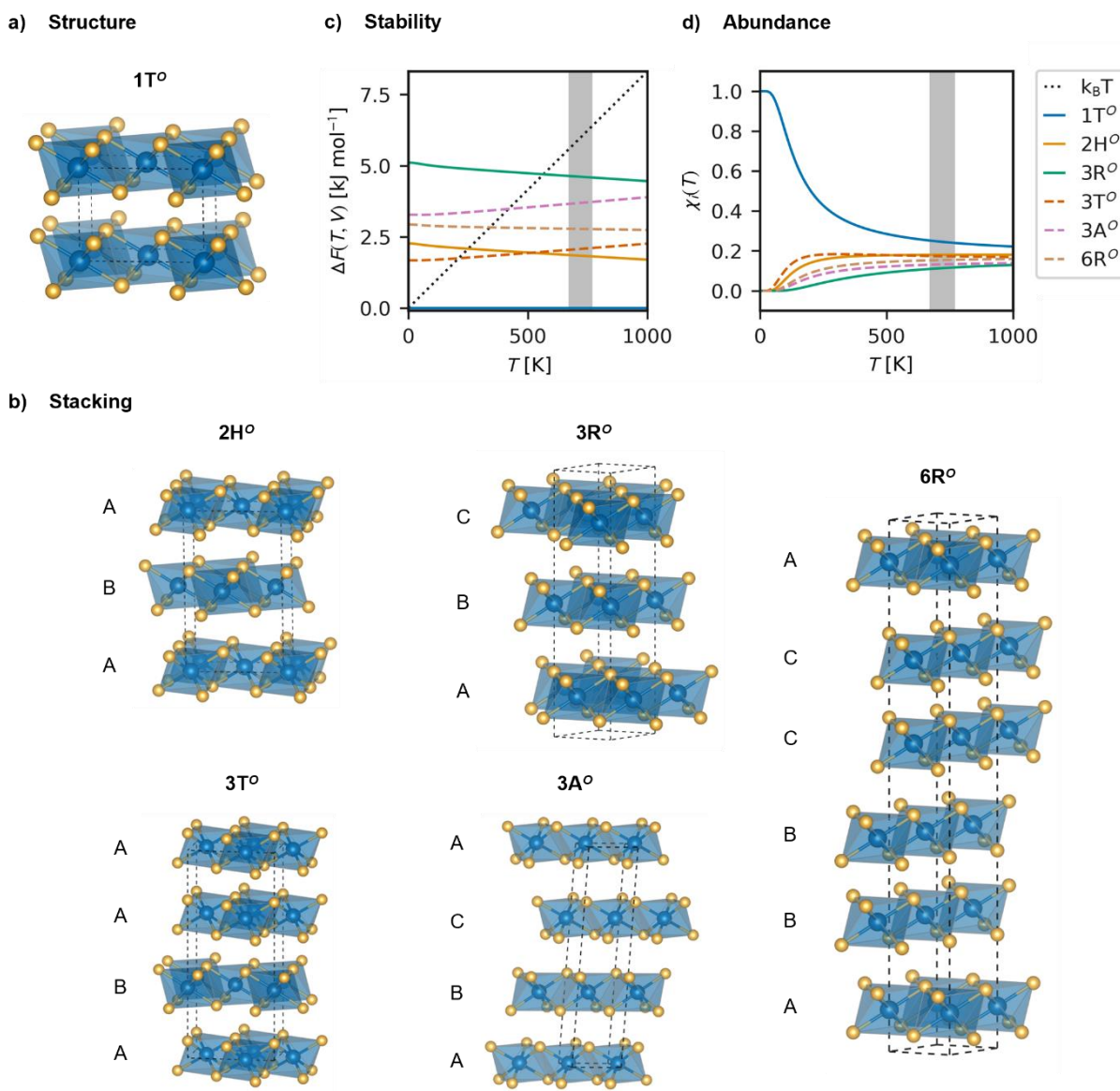

**Figure 1.** a) Most common 1T$^O$ structure of PtSe$_2$. b) The five additional stacking phases studied in this work, including their label, and stacking description (additional data can be found in the **Supporting Information**). c) Relative thermodynamic stability of all six stacking phases based on the Free Helmholtz energy at constant volume. d) Relative thermodynamic abundance based on the partition function at different temperatures. The gray stripe indicates experimental synthesis temperatures.

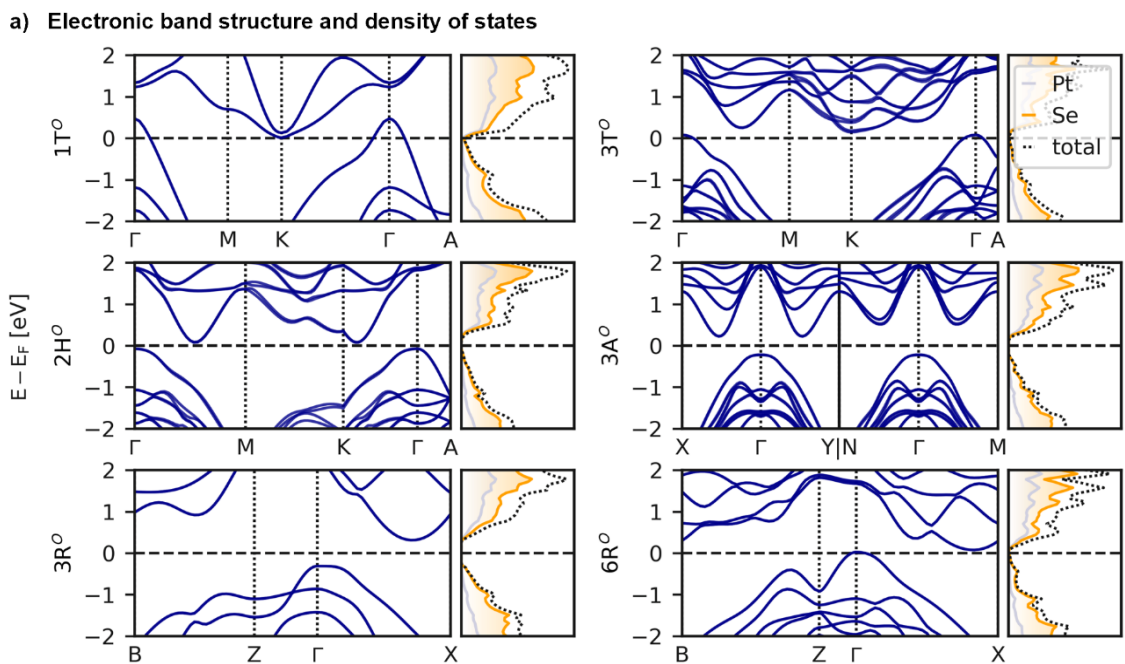

**Figure 2.** a) Electronic band structure and density of states of the six bulk stacking phases at the HSE06 level of theory including spin-orbit coupling (SOC) for a subsection of the Brillouin Zone. The full path along the Brillouin Zone can be found in the **Supporting Information**.

At synthesis temperatures of 400-500 °C,[10] all six stackings become thermodynamically available with their Free Helmholtz Energy differences being lower than $k_B T$ (see **Figure 1c**). Whereas the high-symmetry $1T^O$ and $3R^O$ stacking phases feature no low-lying optical branches in their phonon spectra, the other four phases have states available at excitation energies below 50 cm$^{-1}$ due to the coupling of the acoustic modes of different layers (see **Figure S2**). Thus, the lower-symmetry phases become favored by entropy at elevated temperatures. The $1T^O$ phase is still the most stable one in the temperature range between 0 and 1000 K, but its relative abundance in equilibrium decreases to about 30 % (see **Figure 1c**). The relative abundance shows that at experimental temperatures, all six stacking phases can be present in thin films at relatively evenly distributed fractions, with the $1T^O$ phase being in the majority. The estimation of the abundance does not take external pressure, substrate effects and kinetic effects into account, which may be used to tune the chemical equilibrium in favor of one stacking phase or another. Another important factor is the film thickness, where thin films cannot feature all stacking orders (see **Figure S7**).

Most stackings feature semimetallic characteristics in the bulk (see **Figure 2a**), which should dominate the conductance if grain size and film thickness are sufficiently large and the fraction of the $3R^O$ and $3A^O$ stacking phases is small. These semimetals also feature broad absorption tails ranging into the near-infrared regime (see **Figure 3a**), with the $1T^O$ phase being most prominent. For thin films with a reduced layer number, the semimetallic stacking orders undergo a metal-to-semiconductor transition, whereas the semiconducting phases broaden their band gaps (see **Figure S8**). Consequently, their absorption shifts further into the visible regime. This effect is most pronounced for the $1T^O$ phase, whereas the absorption of the other stackings is affected to much less extent. The experimental absorption spectrum of a 15 nm thick film of $PtSe_2$ is shown in **Figure 3a** for comparison. Due to the $1T^O$ phase being the most abundant, we argue that it dominates the absorbance in a polycrystalline thin film. Hence, the absorption spectrum does not help to distinguish different stacking phases.

Likewise, we conclude that distinguishing stackings by Raman and IR spectroscopy is difficult (see **Figure 3b**). The calculated $E_g$ mode of the $1T^O$ phase is strongly affected by interlayer coupling, shifting from 158.1 cm$^{-1}$ to 172.3 cm$^{-1}$ from bulk to monolayer, respectively (exp.: 174 cm$^{-1}$ for 3 to 5 layers).[20] On the other hand, the $A_{1g}$ mode is less affected by interlayer coupling and shifts very little (exp.: 205 cm$^{-1}$).[20,34] Experimentally, shifts of about 10 cm$^{-1}$ have been observed for the $E_g$ mode of $PtSe_2$ depending on the sample thickness.[20] For thin films, the interlayer coupling of the $E_g$ modes gives rise to a distribution of frequencies, which narrows to a single frequency for thick samples due to stacking rigidity. In **Figure 3b**, we indicate the upper and lower bound of these distributions for each group of modes (e.g., $E_g$, $A_{1g}$, $E_u$ and $A_{1u}$ in the case of the $1T^O$ phase) for up to nine layers, which corresponds to roughly 4.3 nm thickness. The $A_{1u}$ and $E_u$ modes are not Raman-active in bulk and monolayer due to higher symmetry but become symmetry-allowed for few layers, however, they are usually not observed in experiments.[20] For all six stacking phases, their Raman/IR-active modes feature smaller frequency distributions due to weaker interlayer coupling and fall into the variance of the active modes of the $1T^O$ phase. Furthermore, Gulo et al.[34] have shown a large temperature-dependence of the Raman frequencies due to anharmonicity, which can vary by 4 to 6 cm$^{-1}$ at 500 K. Hence, in a polycrystalline mixture, the measurement of Raman shifts alone does not give conclusive evidence of other stacking phases.

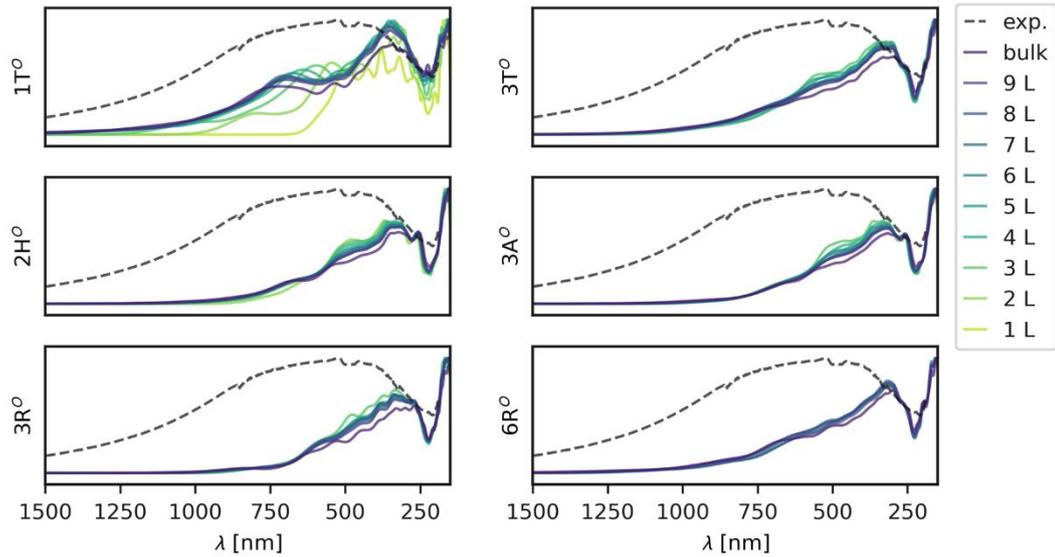

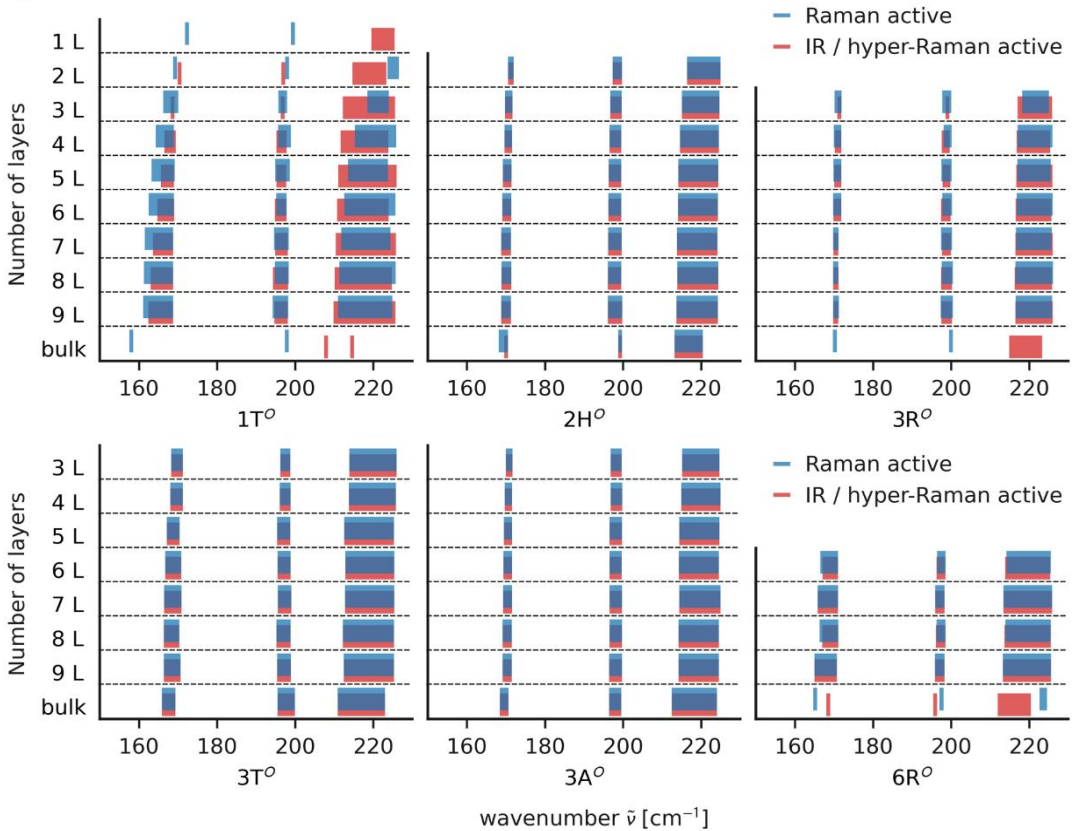

**Figure 3.** a) Calculated absorption spectrum based on the imaginary part of the dielectric function for different layer numbers of the six stacking phases vs. experimental absorption spectrum for a thickness of PtSe$_2$ of 15 nm. b) Frequencies of the Raman- and IR-active modes of the six stacking phases for different layer numbers and bulk. The width of the bar indicates the lower and upper range of the respective modes.

**2.2 Electronic, Thermoelectric and Mechanical Properties**

For nanoelectromechanical applications, the reproducible synthesis of PtSe$_2$ thin films with a large piezoresistive effect is desirable. Polycrystallinity is disadvantageous since the appearance of grain boundaries leads to worse mechanical stability and hinders electrical conductivity. However, it is unclear which stacking phase of PtSe$_2$ is the most desirable one for nanoelectromechanical systems. Film thicknesses between 3-20 nm of PtSe$_2$ are expected to be semimetallic experimentally,[6,18,30] which is beneficial for low contact resistances. On the other hand, semiconductors are expected to have higher gauge factors, such as mono- and bilayer $1T^O$-PtSe$_2$,[6] as well as the $3R^O$ and $3A^O$ stacking phases, and thin films of the $2H^O$, $3T^O$ and $6R^O$ phases (see **Figure S8**).

The relevant quantity for the gauge factor (GF) is the change of the resistivity $\rho$ under strain $\varepsilon$:[1]

$$GF = \frac{\Delta R/R}{\Delta L/L} = \frac{\Delta R/R}{\varepsilon} = 1 + 2\nu + \frac{\Delta\rho/\rho}{\varepsilon}$$

Here, $R$ is the resistance, $L$ is the length and $\nu$ is the Poisson's Ratio. We outline an approach to estimate the GF in the **Supporting Information**, where we distinguish between two possible scenarios of polycrystallinity that are likely for layered structures. In the first case, the samples may feature vertical stacking disorder, but are generally well-aligned in the xy-plane. Then, the biggest contribution to the sample resistivity comes from the in-plane elements of the resistivity tensor and the GF mainly depends on these elements. In the second case, the samples might be randomly aligned with sufficiently large grain sizes, which has been observed in PtSe$_2$ for film thicknesses bigger than 40 to 50 nm.[22] Then, we employ Hill's definition[35] for the Poisson's ratio of a polycrystalline aggregate

$$\nu^* = \frac{1}{2}\left(1 - \frac{3G_V}{3K_V + G_V}\right)$$

with $G_V$ being a mean Voigt shear modulus and $K_V$ being a mean Voigt bulk modulus. In the second case, we average over the trace of the resistivity tensor.

Since the resistivity is a function of both temperature and carrier concentration due to intrinsic doping, so is the GF (see **Figure 4a**). In **Table 1**, we summarize both definitions of the GF at 300 K with the range coming from experimentally observed carrier concentrations. This shows that the carrier concentration must be controlled carefully in experiment, because it may heavily affect the sign and magnitude of the GF. Interestingly, we predict small GFs for the $1T^O$, $2H^O$, $3T^O$ and $3A^O$ stacking phases, whereas the $3R^O$ and $6R^O$ stacking phases feature the largest negative values (up to -370). Hence, these might actually be the most desirable phases for piezoresistive sensing if they can be grown as a majority fraction. Furthermore, the discrepancy between the "in-plane" scenario and the "randomly-aligned" scenario indicates that polycrystallinity may also have a huge impact, with a tendency to worsen the GF. The calculated GFs agree well with experimentally observed GFs (see **Figure 4a**), but do not allow for a distinction of different stacking phases.

This leaves the question how the stacking phases might be assigned in experiment. We argue that a correlation of different techniques to HRTEM or XRD analysis might yield conclusive evidence of the stacking phases, similar to a finger print.[33] Together with electromechanical measurements, the Seebeck coefficient might be a useful indicator for sample quality. We estimate the Seebeck coefficient for a temperature gradient from 300 to 400 K as function of the carrier concentration for different stacking phases (see **Figure 4b** with method details in the

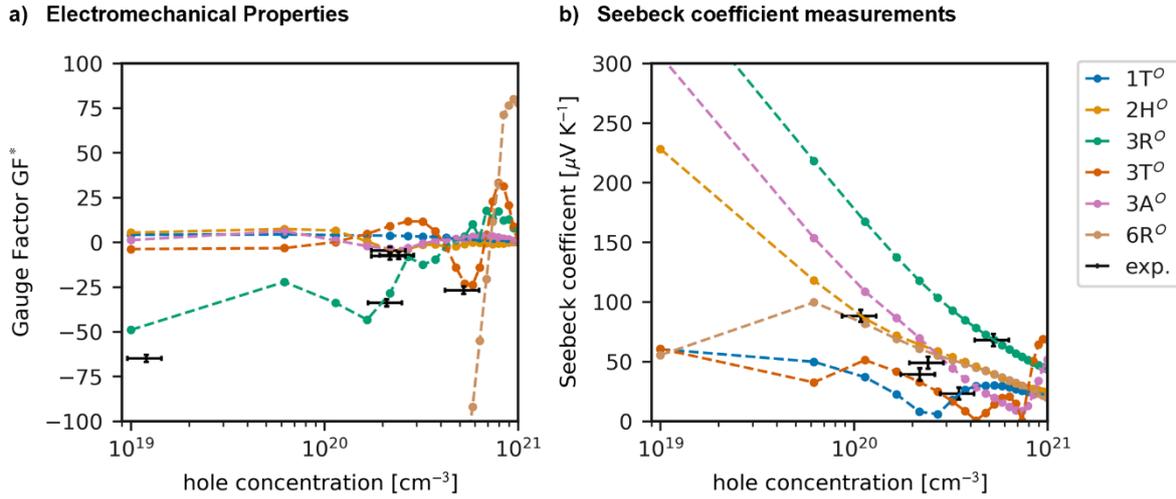

**Figure 4.** a) Experimental gauge factors vs. theoretical estimate of pure phases. b) Experimental Seebeck coefficients vs. theoretical estimate. Bulk carrier concentrations were estimated including an error coming from the film thickness varying between different measurements taken from Lukas et al.[33] and Prechtl et al.[23] and gauge factors were taken from Lukas et al.[33]

**Supporting Information**). The semimetallic stacking phases, such as the $1T^O$ phase, tend to feature smaller Seebeck coefficients. For small carrier concentrations, the Seebeck coefficient is largest with the $3R^O$ phase showing the highest Seebeck coefficient for all stacking phases. The estimated Seebeck coefficients agree well with the experimentally observed range, but do not allow a clear distinction of stacking phases on their own. We propose that a high Seebeck coefficient might indicate a large fraction of the $3R^O$ phase, which could be beneficial for nanoelectromechanical sensing due to its higher GF. On the other hand, we must point out as well that the Seebeck coefficient is affected by other experimental factors, such as the grain size and grain boundaries, which we cannot consider in our theoretical model yet.

## 3. Conclusion

We investigated stacking polymorphism in PtSe$_2$ and found six thermodynamically relevant phases at experimental synthesis temperatures of 400-500 °C. We show that these cannot be distinguished by Raman/IR-spectroscopy or their UV/Vis absorption. Furthermore, we estimate their electromechanical properties and show a huge impact of the stacking order and alignment on the gauge factors, because some of these stacking orders might induce semiconducting properties. We highlight the Seebeck coefficient as a potential indicator of the $3R^O$ phase. We argue that a clear distinction of stacking phases requires a correlative analysis of different measurement techniques and HRTEM or XRD. Such a correlative picture is important to judge the sample quality for piezoresistive sensing, where stacking orders other than the $1T^O$ phase might actually be beneficial.

## 4. Methods

A detailed explanation of the methods can be found in the **Supporting Information**. All calculations were carried using the *Fritz-Haber-Institute ab-intitio materials simulations package* (FHI-aims).[36] The sampling of structures was performed employing the Atomic Simulation Environment (ASE)[37] and the Space Group library (spglib).[38] Phonons and harmonic energies were calculated using FHI-vibes[39] and phonopy.[40] Resistivities and Seebeck coefficients were extracted via Boltzmann Transport Theory as implemented in BoltzTraP2.[41] PtSe$_2$ films were fabricated from sputtered or evaporated platinum (Pt) layers by means of thermally assisted conversion (TAC) as published earlier.[42,43] Measurements of the gauge factors and Seebeck coefficients were undertaken as published earlier.[44]

**Acknowledgements**

This work was financially supported by the German Ministry of Education and Research (BMBF) under the project ForMikro NobleNEMS (16ES1121). We thank the Center for Information Services and High-Performance Computing (ZIH) at TU Dresden for generous allocations of computer time. The authors gratefully acknowledge the Gauss Centre for Supercomputing e.V. (www.gauss-centre.eu) for funding this project by providing computing time through the John von Neumann Institute for Computing (NIC) on the GCS Supercomputer JUWELS at Jülich Supercomputing Centre (JSC). The authors also thank CRC 1415 for support.

We thank Louis Stuber for his technical support and Florian Knoop and Thomas Purcell for fruitful discussions regarding phonon calculations. We thank Jesús Carrete Montaña for fruitful discussions regarding the Seebeck measurement.

**Supporting Information**

**1. Method details**

**1.1. Sampling of stacking phases**

We generated stacking-disordered phases of PtSe$_2$ based on a single-layer of 1T$^O$-PtSe$_2$ by allowing for all possible combinations of five translations and three rotations in up to three layers in a bulk unit cell, resulting in 225 combinations. The allowed translations in fractional coordinates were (0.0 **a** + 0.0 **b**), (0.5 **a** + 0.0 **b**), (1/3 **a** + 1/3 **b**), (1/3 **a** + 2/3 **b**) and (0.5 **a** + 0.5 **b**). The rotation angles considered were 0°, 30° and 60°, which preserve the trigonal lattice. This pool of initial structures was reduced by symmetry analysis as implemented in the Atomic Simulation Environment (ASE)[1] and the Space Group Library (spglib)[2]. The structure (atomic positions and cell parameters) was optimized within a 2 × 2 × 1 supercell to allow for spontaneous symmetry reduction. All calculations were performed using FHI-aims[3] on tight tier 1 numeric atom-centered orbitals employing the PBE functional[4] with added non-local many-body dispersion correction (MBDnl)[5] on Monkhorst-Pack Γ-centered $k$-grids with $k$-point densities of 12 points per Å till the final geometries were converged up to threshold of $5 \cdot 10^{-3}$ eV per Å via FHI-vibes.[6]

**1.2. Calculation of phonons and thermodynamic stability**

For structures remaining from the sampling above, phonons were calculated using phonopy[7] in converged super cells depending on the initial size and shape of the primitive unit (supercell matrices: 2H, 3T, 3A: 2 × 2 × 2; 1T, 6R: 3 × 3 × 3, 3R: [[3,-3,0],[2,2,-4],[1,1,1]]). Structures that exhibited imaginary frequencies were subsequently discarded for being unstable. We calculated the thermodynamic stability with the Free Helmholtz energy: $\Delta F(T,V) = ZPE + F_{phon}(T,V) + E_{elec}(0\ K)$, where the electronic energy was taken from DFT at the PBE level including the MBDnl energy. The Raman/IR-active Γ-point frequencies were extracted with the Bilbao crystallographic server.[8]

**1.3. Calculation of relative abundances**

We estimate relative abundances at experimental temperatures from the partition functions of the bulk solids. Formally, the phononic part of the partition function of a bulk solid is defined as[9]

$$Z_{phon} = \prod_{\mathbf{k},\nu} \frac{e^{-h\nu(\mathbf{k})\cdot 2\beta}}{1 - e^{-h\nu(\mathbf{k})\cdot \beta}}$$

with $\beta$ being the Boltzmann factor. However, the calculation of $Z_{phon}$ is impractical because it diverges at the Γ-point due to the acoustic modes. Still, the Free Helmholtz energy at constant volume can be obtained from the natural logarithm of $Z_{phon}$:

$$F_{phon}(T,V) = -\beta \cdot \ln(Z_{phon})$$

This expression can be simplified by replacing the summations in $\ln(Z_{phon})$ with an integral over the phonon density of states $g(\nu)$:[6,9]

$$F_{phon}(T,V) \approx \int g(\nu)\left(\frac{h\nu}{2} + \beta \cdot \ln(1 - e^{-h\nu\cdot\beta})\right) d\nu$$

This approach is numerically stable because of the vanishing density of states of the acoustic modes at Γ for low frequencies. We use this to estimate $Z_{phon}$:

$$Z_{phon} \approx e^{-F_{phon}(T,V)\cdot\beta}$$

The total partition function is then estimated as $Z_{phon} \cdot Z_{elec}$ and the relative thermodynamic abundance at equilibrium for infinite time is calculated assuming non-interacting phases:

$$\chi_i(T) = \frac{Z_{i,phon} Z_{i,el}}{\sum_j Z_{j,phon} Z_{j,el}} = \frac{Z_{i,phon} \cdot e^{-\Delta E_{i,elec}\cdot\beta}}{\sum_j Z_{j,phon} \cdot e^{-\Delta E_{j,elec}\cdot\beta}}$$

### 1.4. Calculation of electronic and mechanical properties

For the remaining six stacking phases, as well as two-dimensional stacks of these phases for up to nine layers (starting from the smallest number of layers needed to build that stacking order), we calculated the electronic band structures, densities of states, and absorption spectra at the HSE06[10] level including atomistic scalar-relativistic corrections (ZORA) and spin-orbit coupling (SOC) as implemented in FHI-aims on tight tier 1 numeric atom-centered orbitals with k-point line densities of 12 points per Å.[3]

The elastic tensor of the six bulk phases was calculated by numerical forward differentiation of the strained unit cell for a strain of 0.001 Å with subsequent relaxation of atomic positions. The Poisson's ratios were calculated from the compliance tensor as implemented in the matscipy library.[11]

We calculated the Boltzmann conductivity tensor $\boldsymbol{\sigma}/\tau$ as implemented in BoltzTraP2[12] in the constant relaxation time approximation for all bulk stacking phases, as well as for strained systems. We estimate experimental Seebeck coefficients via

$$U \approx -(T_2 - T_1) \cdot \mathbf{Tr}(\mathbf{S}(\bar{T})) \text{ ,[13]}$$

with $\bar{T}$ being the average temperature $(T_1 + T_2)/2$ and $\mathbf{Tr}(\mathbf{S}) = (S_{xx} + S_{yy} + S_{zz})/3$ being the mean trace of the Seebeck tensor. The voltage $U$ is typically plotted as $|\Delta U|/\Delta T$ and the Seebeck coefficient is extracted from the slope via linear regression.[13]

In the following, we motivate two estimates of the gauge factor assuming either that a) the layers are well-aligned in the xy plane or b) randomly aligned and polycrystalline. In both cases, we assume the same uniform relaxation time $\tau$ for the strained and unstrained systems, which allows us to calculate the change of resistivities from the Boltzmann conductivities given relative to an unknown relaxation time:

$$\boldsymbol{\rho}/\tau = \boldsymbol{\sigma}^{-1}/\tau$$

$$a) \quad GF \approx 1 + \nu_{xy} + \nu_{yx} + \frac{1}{2}\left(\frac{\frac{\rho'_{xx} - \rho_{xx}}{\rho_{xx}}}{\varepsilon_{xx}} + \frac{\frac{\rho'_{yy} - \rho_{yy}}{\rho_{yy}}}{\varepsilon_{yy}}\right)$$

Here, $\boldsymbol{\rho}$ is the Boltzmann resistivity tensor, $\boldsymbol{\sigma}$ the Boltzmann conductivity tensor, $\nu_{xy}$ and $\nu_{yx}$ are Poisson's ratios, and $\varepsilon$ is the strain along a certain direction. In case a) we average over the in-plane elements of the Poisson's ratio and the resistivity tensor.

In case b), we employ the definition of polycrystalline elasticity given by Hill[14] to calculate the Poisson's ratio from the Voigt bulk modulus $K_V$ and Voigt shear modulus $G_V$:

$$K_V = \frac{(c_{11} + c_{22} + c_{33}) + 2 \cdot (c_{12} + c_{23} + c_{31})}{9}$$

$$G_V = \frac{(c_{11} + c_{22} + c_{33}) - (c_{12} + c_{23} + c_{31}) + 3 \cdot (c_{44} + c_{55} + c_{66})}{15}$$

Where $c_{ij}$ are elements of the compliance tensor $\mathbf{C}$ in Voigt notation. Then, the polycrystalline Poisson's ratio $\nu^*$ is given by:

$$\nu^* = \frac{1}{2}\left(1 - \frac{3G_V}{3K_V + G_V}\right)$$

And the polycrystalline gauge factor for strain in x-direction GF$^*$ is estimated from the trace average of the resistivity tensor $\rho^* = (\rho_{xx} + \rho_{yy} + \rho_{zz})/3$:

$$b) \quad GF^* \approx 1 + 2 \cdot \nu^* + \left(\frac{\Delta\rho^*}{\rho^*}\right)/\varepsilon_{xx}$$

## 1.5. Experimental methods

PtSe$_2$ films were fabricated from sputtered or evaporated platinum (Pt) layers by means of thermally assisted conversion (TAC) as published earlier.[15,16] To determine the charge carrier concentrations, the PtSe$_2$ films were transferred from their centimeter-scale SiO$_2$ or quartz growth substrates onto highly p-doped Si / 90 nm SiO$_2$ substrates. Six-port Hall devices were defined by optical lithography and reactive ion etching as described in our previous study.[17] Hall measurements were then performed to extract the sheet charge carrier density, as also described in the same study.[17]

For gauge factor (GF) and Seebeck coefficient measurements, the PtSe$_2$ films were transferred onto flexible polyimide substrates and contacted with Nickel (Ni) electrodes, as again described in the same study,[17] including details of the GF measurement using a steel beam set-up. The same samples were used for the measurement of the Seebeck coefficient in a thermoelectric measurement set-up where the two ends of the device were clamped onto hotplates of different temperatures to create a temperature gradient. While one hotplate was kept at room temperature, the other one was heated up to 100 °C gradually. During this process, the voltage between the two Ni contacts of the device was measured using a nanovoltmeter. A linear fit of the recorded voltage versus the temperature difference was used to extract an approximate Seebeck coefficient for the given temperature range.

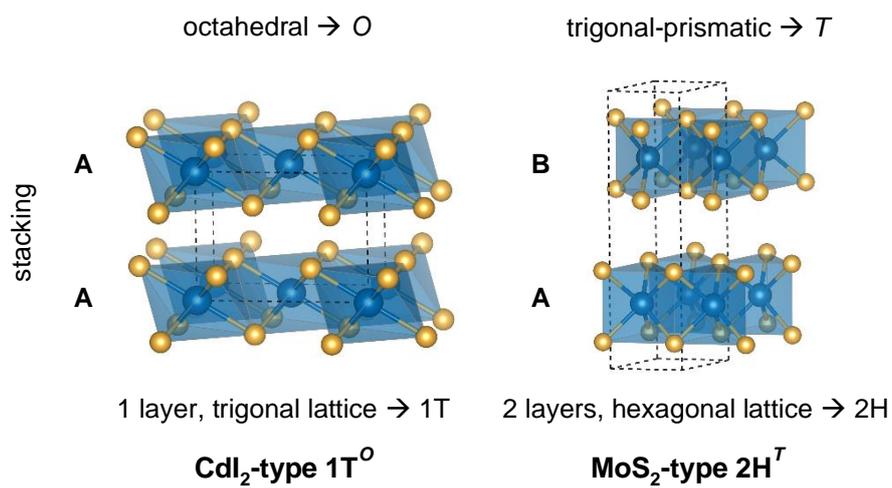

**Figure S1** – Short-hand nomenclature in this work to distinguish between common TMDC polytypes.

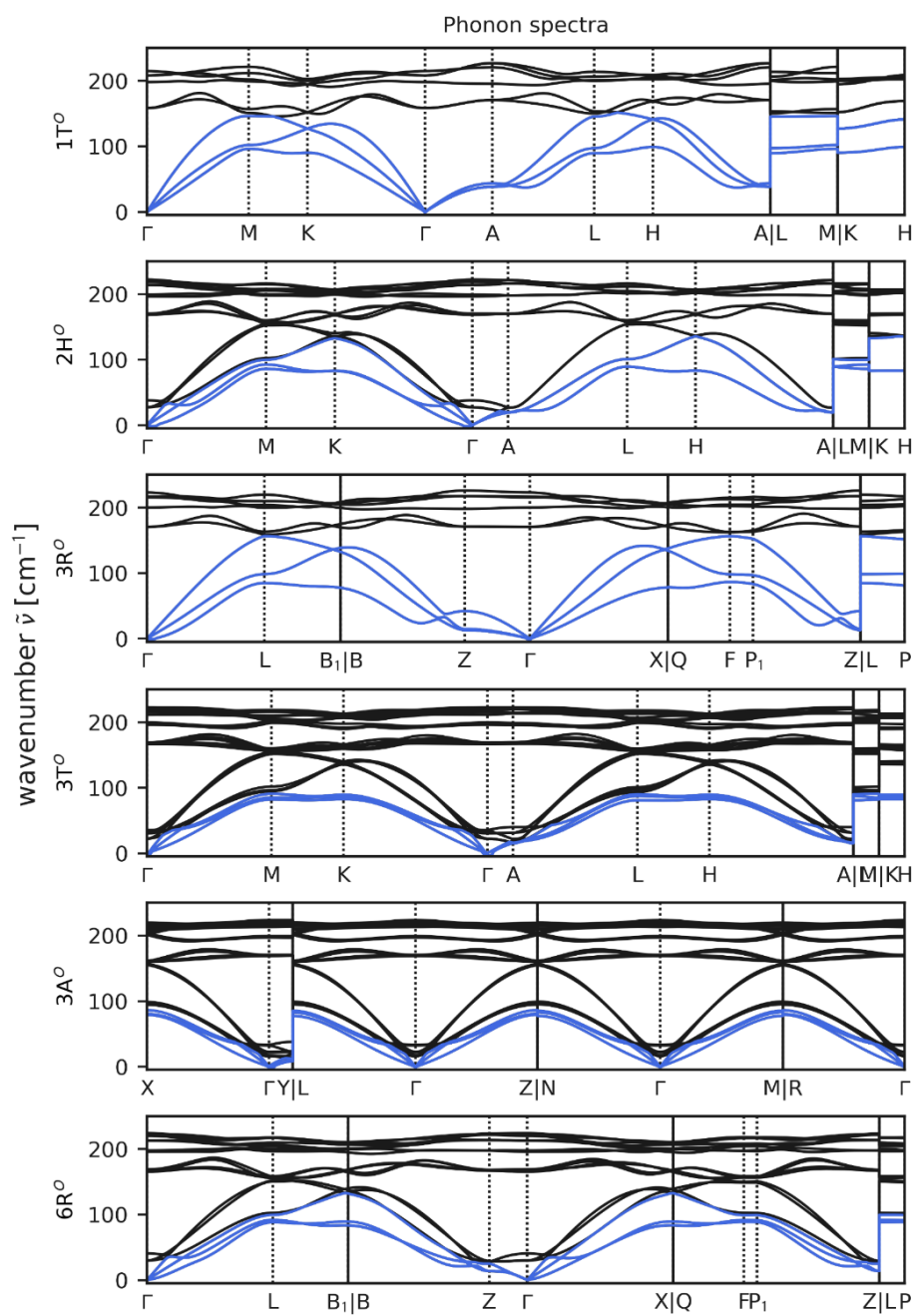

**Figure S2** – Phonon spectra of six stacking phases of PtSe$_2$. The spectra show no imaginary frequencies and the structures are considered locally stable. Blue bands indicate the acoustic modes. The Brillouin zone paths follow the convention of Setyawan-Curtarolo.[1]

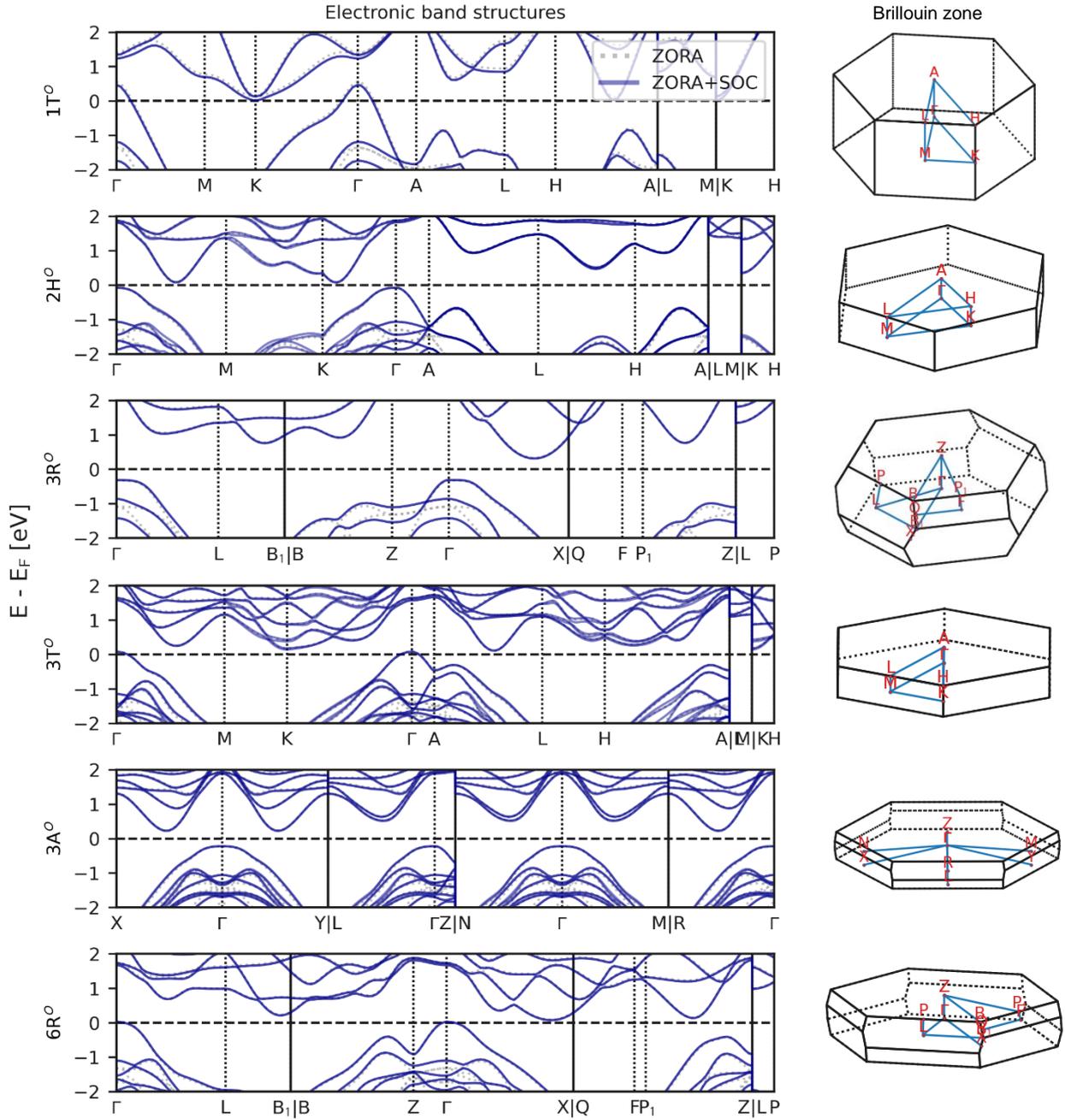

**Figure S3** – Electronic band structures of six stacking phases of PtSe$_2$ at the HSE06 level of theory with and without spin-orbit coupling (SOC). The Brillouin zone paths follow the convention of Setyawan-Curtarolo.[1]

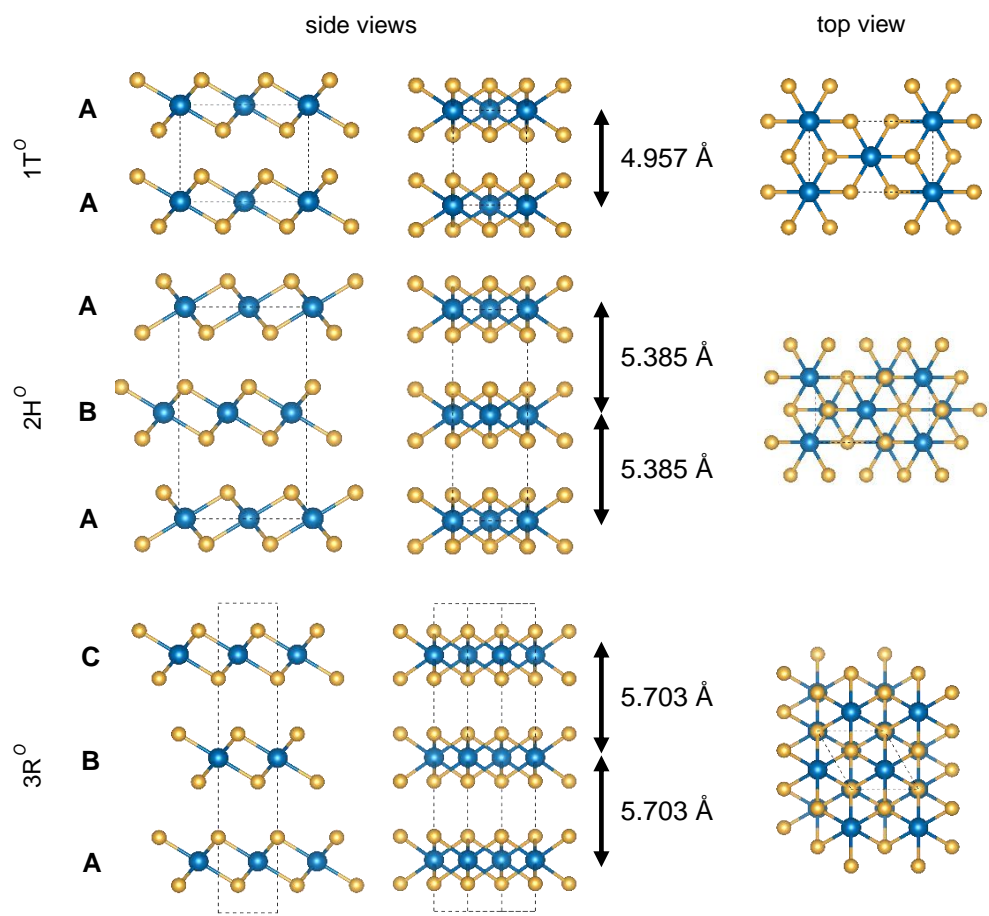

**Figure S4** – Visualization of the 1T$^O$, 2H$^O$ and 3R$^O$ stacking phases from different perspectives including stacking order and interlayer distance.

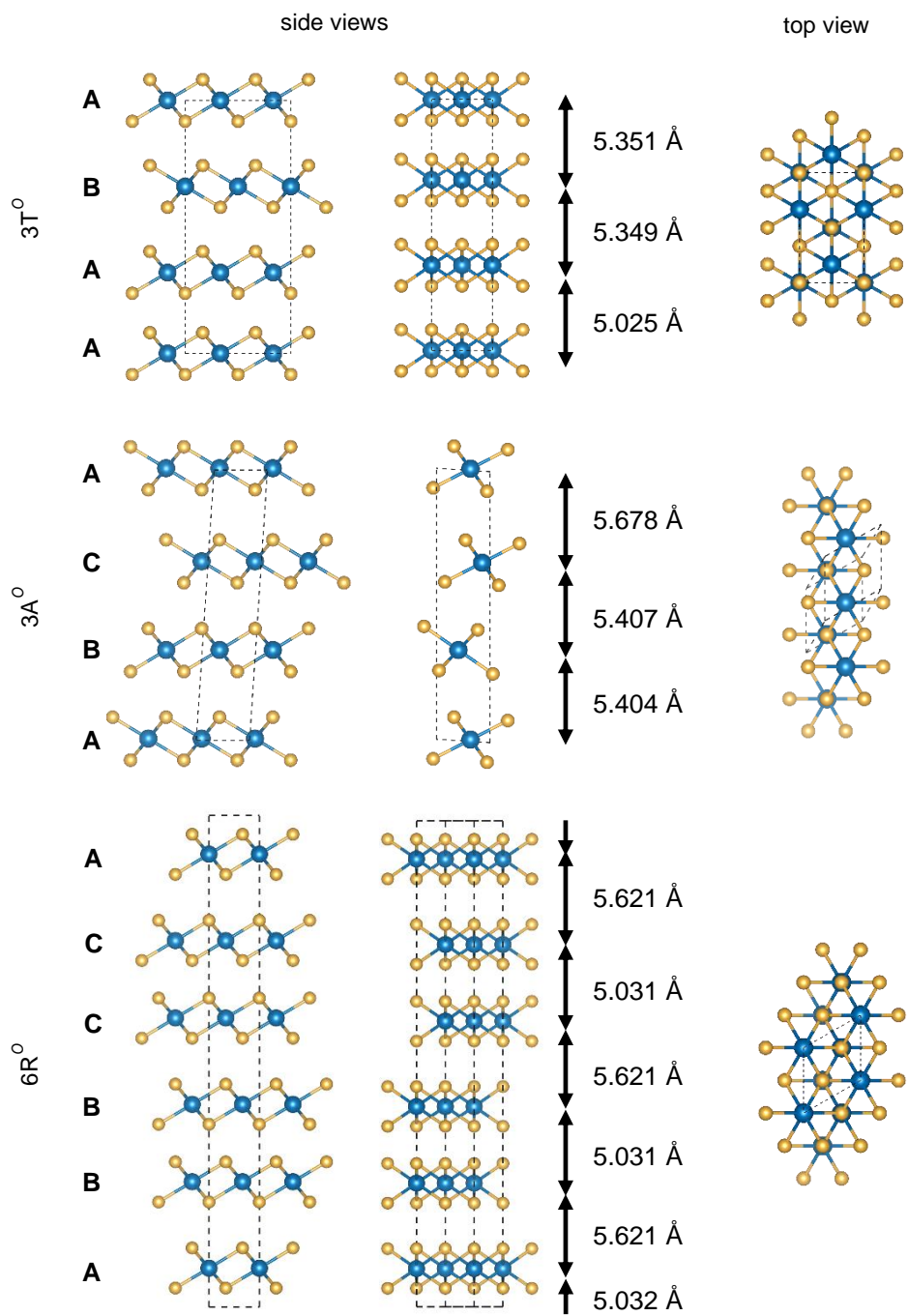

**Figure S5** – Visualization of the 3R$^O$, 3A$^O$ and 6R$^O$ stacking phases from different perspectives including stacking order and interlayer distance.

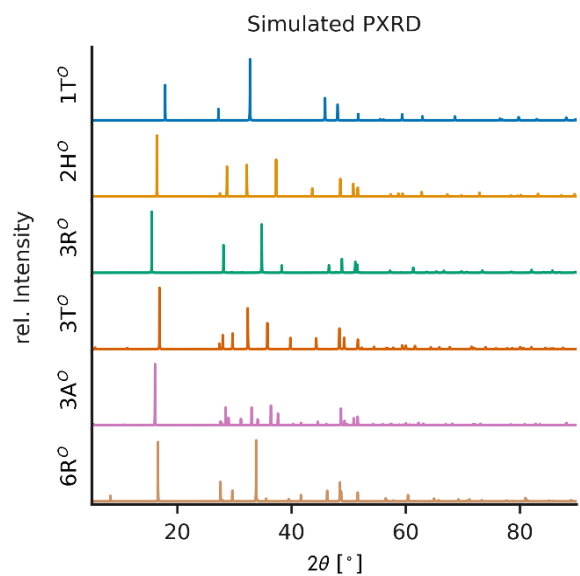

**Figure S6** – Simulated powder x-ray diffraction (PXRD) pattern for the six bulk stacking phases with VESTA for a wavelength of $\lambda = 1.54059$ Å.

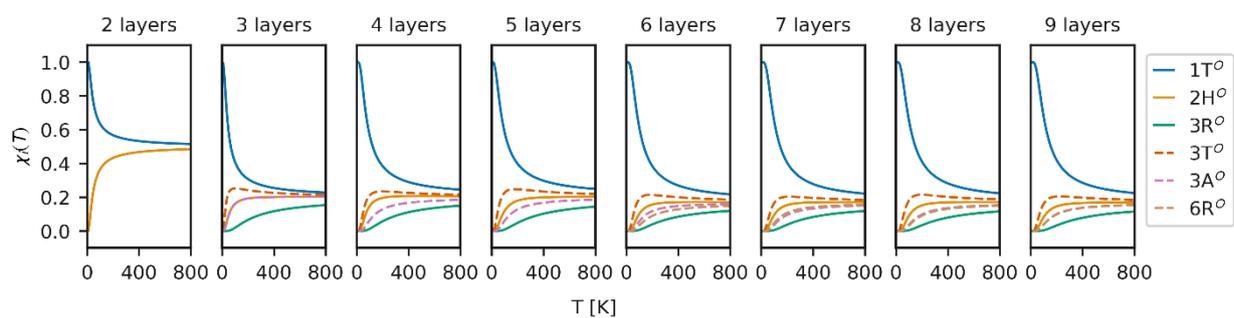

**Figure S7** – Relative abundance vs. temperature for different layer numbers.

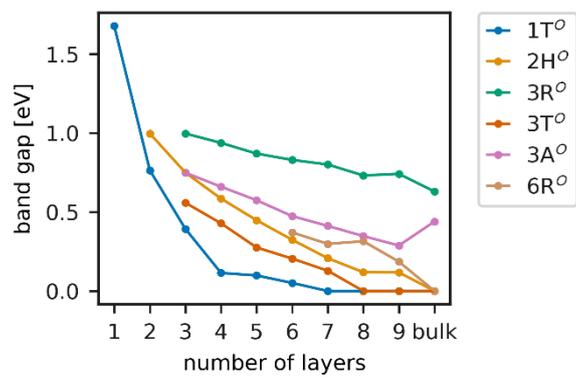

**Figure S8** – Electronic band gap per layer number and stacking phase at the HSE06+SOC level of theory.